\documentclass[preprint,5p,times,twocolumn]{elsarticle}

\usepackage{amssymb}
\usepackage{amsthm}

\usepackage{lineno}
\usepackage{hyperref}
\usepackage{amsmath}
\usepackage[mathscr]{eucal} 
\usepackage[flushleft]{threeparttable} 
\usepackage{algorithm}
\usepackage{algpseudocode}
\usepackage{color}
\usepackage{soul}

\newcommand{\fder}[2]{\dfrac{\mathrm{d}#1}{\mathrm{d}#2}}

\newcommand{\fpartial}[2]{\dfrac{\partial #1}{\partial #2}}

\journal{Journal of Theoretical Biology}

\begin{document}

\begin{frontmatter}



\title{Global sensitivity analysis informed model reduction and selection applied to a Valsalva maneuver model}

\author[um,ncsu]{E. Benjamin Randall\corref{cor1}}
\ead{ebrandal@umich.edu}

\author[unc,ncsu]{Nicholas Z. Randolph}
\ead{nzrandol@unc.edu}

\author[ncsu]{Alen Alexanderian}
\ead{aalexan3@ncsu.edu} 

\author[ncsu]{Mette S. Olufsen\corref{cor2}}
\ead{msolufse@ncsu.edu} 

\cortext[cor1]{Corresponding author}
\cortext[cor2]{Principal corresponding author} 

\address[um]{Department of Molecular and Integrative Physiology, University of Michigan, Ann Arbor, MI}
\address[unc]{Department of Bioinformatics and Computational Biology, University of North Carolina at Chapel Hill, Chapel Hill, NC}
\address[ncsu]{Department of Mathematics, North Carolina State University, Raleigh, NC}


%%%%%%%%%%%%%%%%%%%%%%%%%%%%%%%%%%%%%%%%%%%%%%%%%%%%%%%%%%%%%%%%%%%%%%%%%%%%%%%%%%%%%%%%%%%%%%%%%%%%%%%%%
\begin{abstract}
In this study, we develop a methodology for model reduction and selection informed by global sensitivity analysis (GSA) methods. We apply these techniques to a control model that takes systolic blood pressure and thoracic tissue pressure data as inputs and predicts heart rate in response to the Valsalva maneuver (VM). The study compares four GSA methods based on Sobol' indices (SIs) quantifying the parameter influence on the difference between the model output and the heart rate data. The GSA methods include standard scalar SIs determining the average parameter influence over the time interval studied and three time-varying methods analyzing how parameter influence changes over time. The time-varying methods include a new technique, termed limited-memory SIs, predicting parameter influence using a moving window approach. Using the limited-memory SIs, we perform model reduction and selection to analyze the necessity of modeling both the aortic and carotid baroreceptor regions in response to the VM. We compare the original model to systematically reduced models including (i) the aortic and carotid regions, (ii) the aortic region only, and (iii) the carotid region only. Model selection is done quantitatively using the Akaike and Bayesian Information Criteria and qualitatively by comparing the neurological predictions. Results show that it is necessary to incorporate both the aortic and carotid regions to model the VM. 

\end{abstract}


%%%%%%%%%%%%%%%%%%%%%%%%%%%%%%%%%%%%%%%%%%%%%%%%%%%%%%%%%%%%%%%%%%%%%%%%%%%%%%%%%%%%%%%%%%%%%%%%%%%%%%%%%Research highlights
%	\item Research highlight 1
%	\item Research highlight 2

%%%%%%%%%%%%%%%%%%%%%%%%%%%%%%%%%%%%%%%%%%%%%%%%%%%%%%%%%%%%%%%%%%%%%%%%%%%%%%%%%%%%%%%%%%%%%%%%%%%%%%%%%

\end{frontmatter}



%%%%%%%%%%%%%%%%%%%%%%%%%%%%%%%%%%%%%%%%%%%%%%%%%%%%%%%%%%%%%%%%%%%%%%%%%%%%%%%%%%%%%%%%%%%%%%%%%%%%%%%%%
\pagebreak 

\section*{Abbreviations (alphabetically)} 
\label{Abbreviations}

\begin{tabular}{ll}
	Akaike information criterion with correction & AICc \\ 
	Bayesian information criterion & BIC \\ 
	Delay differential equations & DDE \\ 
	Electrocardiogram & ECG \\ 
	Generalized Sobol' indices & GSIs\\  
	Global sensitivity analysis & GSA \\ 	
	Initial conditions & ICs \\
	Intrathoracic pressure & ITP \\	
	Limited-memory Sobol' indices & LMSIs \\ 
	Local sensitivity analysis & LSA \\ 
	Ordinary differential equations & ODEs \\ 
	Piecewise cubic Hermite interpolating polynomial & PCHIP \\ 
	Pointwise-in-time Sobol' indices & PTSIs \\	
	Quantity of interest & QoI \\ 	
	Respiratory sinus arrhythmia & RSA \\ 						
	Sobol' indices & SIs \\ 
	Systolic blood pressure & SBP \\ 
	Valsalva maneuver & VM \\
\end{tabular}

%%%%%%%%%%%%%%%%%%%%%%%%%%%%%%%%%%%%%%%%%%%%%%%%%%%%%%%%%%%%%%%%%%%%%%%%%%%%%%%%%%%%%%%%%%%%%%%%%%%%%%%%%
\pagebreak

\section{Introduction}
\label{Introduction}

\par \noindent Mathematical models describing cardiovascular processes are typically complex with nonlinear interactions. They have many interrelated states and a large number of parameters. Reducing these models may help model-based data analysis provided that the reduced model retains similar behavior to the original. Several formal model reduction methods exist \cite{Besselink2013} but most focus on analyzing input-output relationships, ignoring predicted quantities. Moreover, they typically identify one reduced model when in fact many can arise. We develop a systematic approach for model reduction and selection via global sensitivity analysis. We then apply this protocol to our model \cite{Randall2019_JAP} that takes systolic blood pressure (SBP) and intrathoracic tissue pressure (ITP) as inputs to predict the heart rate response to the Valsalva maneuver (VM), characterized by forceful exhalation against a resistance. Moreover, we use model selection to investigate whether differentiating between the aortic and carotid baroreceptor afferent signals is necessary for accurate prediction of heart rate. 

\par To analyze parameter influence on a quantity of interest (QoI), we employ {\em local} (LSA) or {\em global} (GSA) sensitivity analysis. LSA (e.g., \cite{Ellwein2008,Marquis2018,Olufsen2013}) computes partial derivatives of the QoI with respect to the nominal parameter values, whereas GSA \cite{Smith2014} computes parameter influence by analyzing parameters and their interactions over the entire parameter space. Here, ``nominal" refers to the {\em a priori} guess and/or calculation used to compute the initial model prediction (before parameter estimation). Popular methods include Sobol' indices (SIs) \cite{Saltelli2010,Sobol2001} (used here), Morris screening \cite{Morris1991,Olsen2019}, generalized sensitivity functions \cite{Kappel2017}, and moment-independent importance measures \cite{Iooss2015}. Morris screening is computationally inexpensive but does not account for higher order interactions. Generalized sensitivity functions characterize model sensitivity to nominal parameter values and take into account parameter interactions, yet make stringent assumptions on local parameter identifiability. Moment-independent importance measures focus on constructing probability density functions that are computationally expensive and intractable for high dimensional systems.

\begin{figure}[!t]
	\centering 
	\includegraphics[scale=0.6]{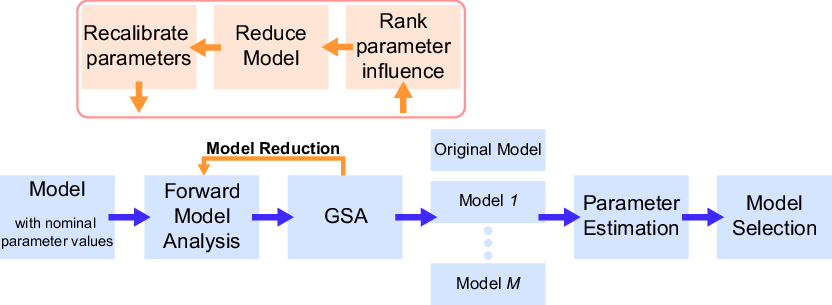}
	\caption{Workflow diagram illustrating the procedure outlined in Section \ref{Methods}. From left to right: A model is developed and nominal parameter values, i.e., the values that provide the initial model prediction, are determined. Forward model analysis produces outputs which undergo global sensitivity analysis (GSA). The GSA results are used to reduce the model (orange arrow), producing $M$ reduced models. For each model, a subset of parameters is estimated to fit data. The reduced model that captures the original model best, qualitatively and quantitatively, is selected. {\it Model Reduction Step} (insert): Using GSA, parameters are ranked from most to least influential, and parameters below a preset threshold are termed noninfluential. These parameters are ``removed" by fixing them at their nominal values or analytically excising the model components associated with them. For each reduced model, nominal parameter values are recalibrated and estimated, and GSA is performed again.}
	\label{workflow}
\end{figure} 

\par We compute our GSA using SIs due to their wide usage and applicability \cite{Calvo2018,Kiparissides2009,Link2018,Sumner2012}. SIs apportion relative contributions of the effect on the QoI to each parameter. Originally developed to analyze parameter influence in models with scalar outputs \cite{Sobol2001} ({\em scalar} SIs), they have recently been extended to analyze time-varying QoIs \cite{Alexanderian2020}. {\em Pointwise-in-time} Sobol' indices (PTSIs) calculate the SI at every time point but do not account for the changes in the pointwise variance over time \cite{Alexanderian2012,Kiparissides2009}. To incorporate time dependence, {\em generalized} Sobol' indices (GSIs) integrate the pointwise variance over a specified time interval \cite{Alexanderian2020,Gamboa2014}. However, for processes that involve significant, short-lived disturbances to the steady-state behavior, integration over the entire time interval diminishes the effect of features that play a transient role. We illustrate this point using our VM model \cite{Randall2019_JAP}, where the breath hold causes a significant and fast model response. To mitigate this issue, we introduce {\em limited-memory} Sobol' indices (LMSIs) that account for parameter interactions and the time history of the variance by applying weights from a moving window. 

\par We also propose a methodology for model reduction and selection. For model reduction, we use the LMSIs to select parameters that are noninfluential over the time interval. These parameters are ``removed" by (i) excision of the equation(s) associated with them \cite{Ellwein2008,Marino2008} or (ii) fixing the them at their nominal values \cite{Sobol2001}. This process generates a set of reduced models upon which we perform model selection using the Akaike Information Criterion with correction (AICc) and Bayesian Information Criterion (BIC) \cite{Wit2012}. Finally, we determine if the reduced models can predict heart rate and autonomic VM responses similar to the original model. 

\par The new contributions of this article include: (i) the development of LMSIs for time-dependent processes to analyze our model's \cite{Randall2019_JAP} response to the VM; and (ii) a GSA-informed model reduction and selection protocol. We use model selection to test whether it is necessary to include afferent signaling from the carotid region only, the aortic region only, or both regions simultaneously to predict heart rate dynamics and neural tones.

%%%%%%%%%%%%%%%%%%%%%%%%%%%%%%%%%%%%%%%%%%%%%%%%%%%%%%%%%%%%%%%%%%%%%%%%%%%%%%%%%%%%%%%%%%%%%%%%%%%%%%%%%
\section{Methods}
\label{Methods}

\par \noindent  Figure \ref{workflow} depicts the model reduction and selection workflow, which includes the following components:

\begin{enumerate} 
	\item {\it Forward model analysis}: We consider the QoI, which is the time-varying residual vector 
	\begin{equation}
		\mathbf{r}(t_j;\theta) = \frac{H(t_j;\theta)-H_d(t_j)}{H_d(t_j)} 
		\label{residual}
	\end{equation}
	where $H(t_j;\theta)$ denotes the heart rate model output at time $t_j$ for $j = 1,\dots,N$ and $H_d(t_j)$ the corresponding heart rate data. Model predictions depend on the parameter vector $\theta \in \Omega_p \subseteq \mathbb{R}^p$, for $\Omega_p$ the parameter space of dimension $p$. 
	
	\item {\it Global sensitivity analysis}: To determine parameter influence on the QoI, we compute scalar SIs with respect to
	\begin{equation} 
		||\mathbf{r}(t;\theta)||_2 = \Bigg( \sum_{j = 1}^{N} \mathbf{r}(t_j;\theta)^2 \Bigg)^{1/2}
		\label{normr}
	\end{equation} 
	 as well as the time-varying PTSIs, GSIs, and LMSIs with respect to $\mathbf{r}$.  
	\begin{enumerate} 
		\item {\it Parameter influence} is computed with the chosen index. We compute LMSIs to determine noninfluential parameters having an index below a given threshold. 
	
		\item {\it Model reduction} ``removes" noninfluential parameters by fixing them at their nominal values or analytically removing equations associated with them. 
		
		\item {\it Model recalibration} ensures that the reduced model produces similar predictions as the original. 	
		
		\item {\it GSA on the reduced model} is conducted to test if all parameters are influential. This step is necessary since model reduction can change the relative ranking of parameter influence and result in some parameters becoming noninfluential. 
	\end{enumerate}
	Steps (b)-(d) are repeated iteratively until the system has no noninfluential parameters. At any point, there may be cases where one could remove one parameter over another, creating branches of reduced models. 
	
	\item {\it Reduced models}: We create a set $\mathscr{M} = \{m_0, m_1, \dots, m_M\}$ for $m_0$ the original model and $m_k$ the reduced models for $k=1,\dots,M$. 
	
	\item {\it Parameter estimation}: To determine an identifiable parameter subset ($\hat{\theta}$), we use subset selection \cite{Marino2008,Olsen2019,Olufsen2013,Pope2009}, which is local in nature, since the model is evaluated at specific parameter values. Each reduced model is fitted to data, estimating $\hat{\theta}$ by minimizing the least squares error. 
	
	\item {\it Model selection}: To ensure that the reduced models produce outputs within physiological ranges, we perform model selection both quantitatively and qualitatively. Quantitatively, we analyze goodness of fit by calculating relative AICc and BIC values. Qualitatively, we assume the original model is the true signal and analyze the predicted neural tones of the reduced models. 
\end{enumerate}

\begin{figure}[!t]
	\centering 
	\includegraphics[]{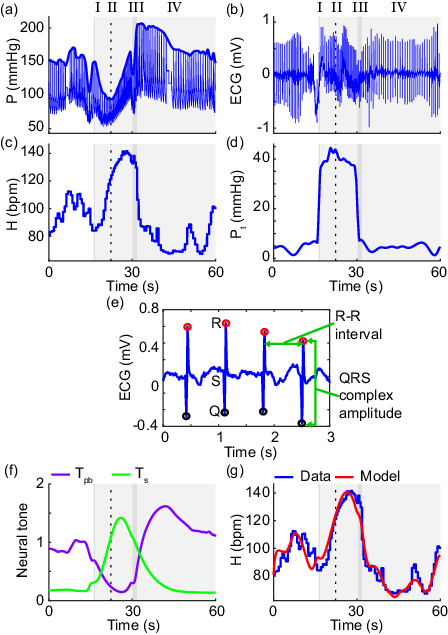}
	\caption{Valsalva maneuver (VM) for a representative subject. The VM phases are represented as alternating gray (I and III) and light gray (II and IV) boxes. Phase II is divided into early and late stages (vertical dotted black line). {\em Phases}: I. Initiation of the breath hold causes a sharp increase in blood pressure (P, mmHg) and a decrease in heart rate (H, bpm). Early II. P drops and H accelerates. Late II. H accelerates further, which increases P. III. Release of the breath hold causes a sharp drop in P. IV. P overshoots and H undershoots, both returning to baseline within 30 s. (a) P with systolic blood pressure (bold) indicated. (b) Electrocardiogram (ECG, mV). (c) H. (d) Thoracic pressure ($P_{t}$, mmHg, equation \eqref{Pth}). (e) ECG zoom with Q (black circles), R (red circles), and S waves indicated. Green arrows show the amplitude of the QRS-complex  (the set of Q, R, and S waves) and the R-R interval (the time between two consecutive R waves). (f) Baroreflex-mediated parasympathetic ($T_{pb}$, dimensionless, purple) and sympathetic ($T_s$, dimensionless, green) neural tones. (g) Heart rate model output ($H$, bpm, red) fitted to data (blue). Optimized parameter values for panels f and g are listed in Table \ref{parameters}. }
	\label{data} 
\end{figure} 

\subsection{The Valsalva maneuver} 
\label{Valsalvamaneuver}

\par \noindent The VM is a clinical test involving forced expiration against a closed airway while maintaining an open glottis \cite{Hamilton1944}.  It stimulates the autonomic nervous system via the baroreflex in response to blood pressure changes in the aortic arch and the carotid sinus. The body responds to a sudden decrease in blood pressure, causing compensatory responses in heart rate to restore blood pressure to baseline. Figure \ref{data} displays a representative VM and includes the four VM phases.

\subsection{Data acquisition and processing} 
\label{Dataacquisitionandprocessing}

\par \noindent The data is from a 21-year-old healthy female following our previous work \cite{Randall2019_JAP}. Figure \ref{data} displays the electrocardiogram (ECG), blood pressure, and heart rate. SBP (Figure \ref{data}a) is obtained by interpolating the local maxima of consecutive blood pressure waveforms using a piecewise cubic Hermite interpolating polynomial (PCHIP). Similarly, we construct a respiration signal from the ECG (Figure \ref{data}b) using PCHIP to detect amplitudes of the QRS complexes (sets of Q, R, and S waves in the ECG) (Figure \ref{data}e). Figure \ref{data}d plots thoracic pressure ($P_t$, equation \eqref{Pth}) during the VM. Heart rate (Figure \ref{data}c) is computed from the ECG by detecting R-R intervals, the time between two consecutive R waves (Figure \ref{data}e).

\begin{figure}[!t]
	\centering
	\includegraphics[scale=0.45]{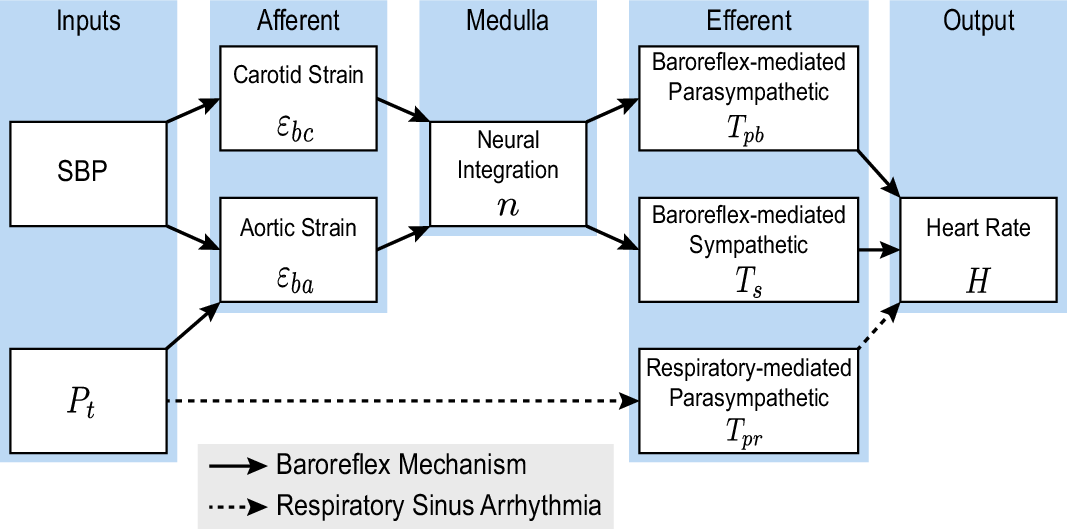}
	\caption{Model diagram reproduced from \cite{Randall2019_JAP}. Systolic blood pressure (SBP, mmHg, Figure \ref{data}a) and thoracic pressure ($P_{t}$, mmHg, Figure \ref{data}d, equation \eqref{Pth}) are inputs. {\em Baroreflex pathway}: (solid arrows) Carotid baroreceptor strain ($\varepsilon_{bc}$, dimensionless, equation \eqref{ebc}) depends on SBP only and aortic strain ($\varepsilon_{ba}$, dimensionless, equation \eqref{eba}) depends on SBP and $P_{t}$. They are integrated in the medulla ($n$, s$^{-1}$, equation \eqref{n}), which initiates a dynamic change in baroreflex-mediated parasympathetic ($T_{pb}$, dimensionless, equation \eqref{Tpb}) and sympathetic ($T_{s}$, dimensionless, equation \eqref{Ts}) tones. {\em Respiratory sinus arrhythmia (RSA) pathway}: (dotted arrows) RSA depends solely on $P_{t}$ that modulates the RSA-mediated parasympathetic tone ($T_{pr}$, dimensionless, equation \eqref{Tpr}). All efferent outflows combine to control heart rate ($H$, bpm, equation \eqref{Hder}).}
	\label{schematic} 
\end{figure}

\begin{table}[!t]
	\centering 
	\footnotesize
	\renewcommand{\arraystretch}{1}
	\begin{threeparttable} 
		\caption{Description of the model parameters and units. Nominal (Nom) values, lower (LB) and upper (UB) bounds, along with an explanation for the chosen bounds. Optimized (Opt) parameters for the model outputs in Figure \ref{data}f and \ref{data}g are listed in the last column.}
		\begin{tabular}{cccccl|c}
			\hline
			\multicolumn{1}{c}{{\bf Symbol}} &  {\bf Units} &  {\bf Nom} & {\bf LB} & {\bf UB} & \multicolumn{1}{c}{{\bf Explanation}} & {\bf Opt} \\
			\hline 
			\hline 
			\multicolumn{6}{l|}{{\em Offset}} & \\ 
			$A$ & & 5 & 2.5 & 7.5 & Nom $\pm$ 50\% & \\
			\hline 
			\multicolumn{6}{l|}{{\em Convex parameter}} & \\  
			$B$ & s$^{-1}$ & 0.5 & 0.01 & 1 &$[0, 1]$ & 0.43$^*$  \\ 
			\hline
			\multicolumn{6}{l|}{{\em Afferent and efferent gains}} & \\  
			$K_b$		 & & 0.1 & 0.05 & 0.15 & Nom $\pm$ 50\%  & \\
			$K_{pb}$	& & 5   & 2.5   & 7.5   & Nom $\pm$ 50\%  & \\
			$K_{pr}$ 	 & & 1   & 0.5   & 1.5   & Nom $\pm$ 50\%  & \\
			$K_s$ 		  & & 5   & 2.5   &7.5   & Nom $\pm$ 50\%  & \\
			\hline 
			\multicolumn{6}{l|}{{\em Time-scales}} &  \\
			$\tau_b$ 	  & s & 0.9 & 0.45 & 1.35   & Nom $\pm$ 50\% & \\
			$\tau_{pb}$	& s & 6.5  & 0.01 & 17.9 & Mean $\pm$ 2SD  & 4.26$^*$  \\ 
			$\tau_{pr}$	 & s & 9.6 & 0.01 & 31.2 & Mean $\pm$ 2SD  & 2.75$^*$ \\ 
			$\tau_s$ 	  & s & 10   & 5      & 15    & Nom $\pm$ 50\% &  \\ 
			$\tau_H$ 	 & s & 0.5  & 0.25 & 0.75 & Nom $\pm$ 50\% & \\
			\hline 
			\multicolumn{6}{l|}{{\em Sigmoid steepnesses}} & \\  
			$q_w$ 	    & mmHg$^{-1}$ & 0.04 & 0.02 & 0.06 &Nom $\pm$ 50\%   & \\  
			$q_{pb}$   & s                    & 10     & 5      &15     & Nom $\pm$ 50\%  & \\ 
			$q_{pr}$  	& mmHg$^{-1}$ & 1      & 0.5   & 1.5    & Nom $\pm$ 50\%  & \\ 
			$q_s$ 		 & s                    & 10     & 5     & 15      & Nom $\pm$ 50\%  & \\ 
			\hline 
			\multicolumn{6}{l|}{{\em Half-saturation values }} & \\
			$s_w$ 		& mmHg     & 123   & 83    & 163 & Mean $\pm$ 2SD   & 149$^{**}$\\  
			$s_{pb}$   & s$^{-1}$  & 0.54  & 0.53 & 0.55 & Mean $\pm$ 2SD & 0.54$^{**}$\\ 
			$s_{pr}$    & mmHg     & 4.88  & 4.46 & 5.3 & Mean $\pm$ 2SD   & 4.50$^{**}$\\ 
			$s_s$        & s$^{-1}$  & 0.05  & 0.04 & 0.06 & Mean $\pm$ 2SD & \\
			\hline  
			\multicolumn{6}{l|}{{\em Heart rate gains }} & \\ 
			$H_I$ 	 	 & bpm  & 100  & 86 & 114  & Mean $\pm$ 2SD  & 106$^{**}$ \\ 
			$H_{pb}$  &          & 0.5  & 0.1 & 0.9  & Mean $\pm$ 2SD  & 0.44$^*$  \\ 
			$H_{pr}$   &          & 0.3  & 0.01 & 1.1 & Mean $\pm$ 2SD  & 0.48$^*$ \\ 
			$H_s$ 		&          & 0.3 & 0.01 & 1.1 & Mean $\pm$ 2SD  & 0.28$^*$ \\ 
			\hline 
			\multicolumn{6}{l|}{{\em VM start and end }} &  \\
			$t_s$ & s & data & & &  & 16.2$^{**}$   \\
			$t_e$ & s & data & & &  & 30.2$^{**}$  \\
			\hline 
			\multicolumn{6}{l|}{{\em Delay }$^{***}$} & \\  
			$D_s$ & s & 3     & & & & \\
			\hline 
		\end{tabular} 
		\label{parameters}
		\begin{tablenotes} 
			\small
			\item SD - standard deviation. VM - Valsalva maneuver. 
			\item  A blank space in the Units column indicates that the parameter is dimensionless. 
			\item A blank space in the Opt column indicates that the parameter is not estimated, i.e., it is kept at its nominal value. 
			\item $^{*}$ The parameter was optimized to fit the data.  
			\item $^{**}$ The parameter was calculated from data. 
			\item $^{***}$ $D_s$ is included in the model but not analyzed. This parameter is held fixed at its nominal value. 

		\end{tablenotes}
	\end{threeparttable}
	\renewcommand{\arraystretch}{1}  
\end{table}

\subsection{Model development} 
\label{Modeldevelopment}

\par \noindent  The model is a system of stiff ordinary (ODEs) and delay (DDEs) differential equations with 6 states and 25 parameters that takes SBP and ITP as inputs to predict heart rate, $H$, in response to a VM. Figure \ref{schematic} displays the  model schematic with the baroreflex (solid arrows) and respiratory sinus arrhythmia (RSA, dotted arrows) pathways. The model is of the form  
\begin{equation} 
	\fder{\mathbf{x}(t)}{t} = f(t, \mathbf{x}(t), \mathbf{x}(t - D_s); \theta), \quad \mathbf{x}(t) = \mathbf{x}_0 \ \text{for} \ t\in[-D_s,0].
	\label{summary}
\end{equation}
Here, $\mathbf{x}(t) = [\varepsilon_{bc}(t), \varepsilon_{ba}(t), T_{pb}(t), T_{pr}(t), T_s(t),H(t)]^T \in \mathbb{R}^6$, where $\varepsilon_{bc}$ and $\varepsilon_{ba}$ are the carotid and aortic baroreceptor strains, respectively, $T_{pb}$ and $T_s$ are the baroreflex-mediated parasympathetic and sympathetic tones, respectively, and $T_{pr}$ is the RSA-mediated parasympathetic tone. $f: \mathbb{R}^{1 + 2(6) + 25} \rightarrow \mathbb{R}^6$ is the right hand side of the system, $\mathbf{x}_0$ is the vector of initial conditions, $D_s \in \mathbb{R}$ (s) is the discrete delay, and $\theta \in \mathbb{R}^{25}$ is a vector of parameters, including 
\begin{multline} 
	\theta = [A, B, K_b, K_{pb}, K_{pr}, K_s, \tau_b,\tau_{pb}, \tau_{pr}, \tau_{s}, \tau_H, q_w, q_{pb}, \\ q_{pr}, q_s, s_w, s_{pb}, s_{pr}, s_s, H_I, H_{pb}, H_{pr}, H_s, t_s, t_e]^T.
	\label{eq:parms}
\end{multline} 
Table \ref{parameters} lists the nominal parameter values. All states except $T_s$ and $H$ are modeled as a first-order linear equation of the form
\begin{equation}
	\fder{x}{t} = \dfrac{-x(t) + K_x G_x(t)}{\tau_x},
	\label{dxdt} 
\end{equation} 
where $x$ is a state, $K_x$ is the gain, $G_x$ is a nonlinear function, and $\tau_x$ is the time-scale. For $\varepsilon_{bc}$ and $\varepsilon_{ba}$, $G_x(t)$ takes the form  
\begin{equation}
	G_{bj}(t) = 1 - \sqrt{ \dfrac{1 + e^{-q_w (P_j(t) - s_w)}}{A + e^{-q_w (P_j(t) - s_w)}}}
	\label{ewj} 
\end{equation} 
for $j = c$ or $a$ for carotid or aortic, respectively, where $P_j$ denotes the pressure input, $A$ (dimensionless) is an offset parameter, and $q_w$ (mmHg$^{-1}$) and $s_w$ (mmHg) are  the steepness and half-saturation values. For $T_{pb}$ and $T_s$, $G_x(t)$ takes the form 
\begin{equation}
	G_{l}(t) = \dfrac{1}{1 + e^{q_{l} (n(t) - s_{l})}} 
	\label{Gl} 
\end{equation}  
for $l = pb$ and $s$ for baroreflex-mediated parasympathetic and sympathetic, respectively, where $q_l$ (s) and $s_l$ (s$^{-1}$) are the steepness and half-saturation values, respectively, and 
\begin{equation}
	n(t) = B (G_{bc} (t) - \varepsilon_{bc}(t) )  + (1 - B) (G_{ba} (t) - \varepsilon_{ba}(t))
	\label{n} 
\end{equation}
for $B \in [0,1]$ (s$^{-1}$). Figure \ref{data}f shows the optimized model predictions for $T_{pb}$ and $T_s$. For $T_{pr}$, $G_x(t)$ takes a similar form as equation \eqref{Gl}, but the time-varying input is $P_t$ (equation \eqref{Pth}). Lastly, $H$ is determined by the ODE 
\begin{equation}  
	\fder{H(t)}{t} = \dfrac{-H(t) + \tilde{H}(t)}{\tau_H}, 
	\label{Hder}
\end{equation} 
where  $\tau_H$ (s) is a time-scale and 
\begin{equation} 
	\tilde{H}(t) = H_I \big(1 - H_{pb} T_{pb}(t) + H_{pr} T_{pr}(t) + H_s T_s (t) \big).
\end{equation} 
$H_I$ (bpm) is the intrinsic heart rate, whereas $H_{pb}$, $H_{pr}$, and $H_s$ are dimensionless gains. The model equations are summarized in \ref{appendix:fulleqs}. Figure \ref{data}g shows the optimized model fit to data.  

\par Initial condition calculations are in \ref{appendix:ICs}. The model is solved using the stiff DDE solver RADAR5 \cite{Guglielmi2001} with relative and absolute tolerances of $10^{-8}$. As described in \ref{appendix:subsetselection}, $H$ was fitted to the heart rate data by estimating a parameter subset ($\hat{\theta}$, equation \eqref{thetahat}) that minimizes the least squares error ($J$, equation \eqref{J}) between the model output and the data. The GSA and parameter estimation are performed on the logarithm of the parameters. We assume that the logarithm of each parameter is uniformly distributed between the logarithm of their respective parameter ranges, given in Table \ref{parameters}. The data and run-time environment can be found in \cite{Randall2020_code}. 

\par Upper and lower parameter bounds are given in Table \ref{parameters}. Parameters not calculated {\it a priori} are varied $\pm$50\% of their nominal values. Parameters calculated from data have bounds set to the mean $\pm$2 SD. To prevent negative parameter values, we set the lower bound to 0.01 where necessary.  The parameter $B$ varies from 0 to 1. The parameter $\tau_s$ is an exception since it interacts with $D_s$ and can cause instability \cite{Randall2019_MathBiosci}. Hence, $\tau_s$ is varied $\pm$50\% of its nominal value to remain stable.  

\subsection{Global sensitivity analysis}
\label{Globalsensitivityanalysis} 

\par \noindent We use variance-based SIs \cite{Saltelli2010,Sobol2001} for the GSA. This section discusses four different methods for computing SIs: scalar \cite{Saltelli2010}, pointwise-in-time \cite{Alexanderian2012}, generalized \cite{Alexanderian2020}, and limited-memory (new). A reference table of symbols is compiled in Table \ref{sobolsymbols}. 

\subsubsection{Scalar Sobol' indices}
\label{ScalarSobolindices}

\par \noindent  Consider a mathematical model $f$ with a scalar output $y$ dependent on $\theta \in \Omega_p \subseteq \mathbb{R}^p$, a random vector of $p$ statistically independent model parameters with a uniform distribution spanning $\Omega_p$; that is, $y= f(\mathbf{\theta})$. For  each parameter $\theta_i$, we compute its contribution to the variance of $y$ \cite{Sobol1993,Sobol2001}. Assuming parameter independence, the main effect on $f$ due to $\theta_i$ is 
\begin{equation} 
	\mathscr{S}_i (f)= \dfrac{\mathbb{V}_{\theta_i} (\mathbb{E}_{\theta_{\sim i}} [f | \theta_i ])}{\mathbb{V}(f)},
	\label{scalarfosobol} 
\end{equation} 
where $\mathbb{V}(\cdot)$ and $\mathbb{E}[\cdot]$ denote the variance and expectation operators and $\theta_{\sim i}$ is the vector $\theta$ without $\theta_i$. The total-effect SI on $f$ includes both the main and higher order effects of $\theta_i$ and is
\begin{equation} 
	\mathscr{T}_i (f)= \dfrac{\mathbb{E}_{\theta_{\sim i}}[\mathbb{V}_{\theta_i}(f |\theta_{\sim i})]}{\mathbb{V}(f)} = 1 - \dfrac{\mathbb{V}_{\theta_{\sim i}} (\mathbb{E}_{\theta_i} [f | \theta_{\sim i}])}{\mathbb{V}(f)}.
	\label{scalartesobol}
\end{equation} 

\begin{table}[!t]
	\centering 
	%\footnotesize
	\renewcommand{\arraystretch}{1} 
	\begin{threeparttable} 
		\caption{Summary of Sobol' index (SI) symbols.}
		\begin{tabular}{lllc}
			\hline
			\multicolumn{1}{c}{{\bf SI Type}} & \multicolumn{2}{c}{{\bf Symbol}}  & {\bf Reference} \\ 
			\cline{2-3}
			& Main Effect & Total Effect &  \\ 
			\hline
			\hline
			Scalar  & $\mathscr{S}_i(f)$ & $\mathscr{T}_i(f)$ & \cite{Saltelli2010}\\
			PTSIs & $\mathscr{S}_i(f; t)$ & $\mathscr{T}_i (f;t)$ & \cite{Alexanderian2012} \\ 
			GSIs & $\mathscr{S}_i(f; [0, t])$ & $\mathscr{T}_i (f;[0, t] )$ &  \cite{Alexanderian2020} \\ 
			LMSIs & $\mathscr{S}_i(f; [t - \Delta, t])$ & $\mathscr{T}_i(f; [t - \Delta ,t] )$ &  This study\\ 
			\hline
		\end{tabular} 
		\label{sobolsymbols}
		\begin{tablenotes} 
			\small
			\item PTSIs - pointwise-in-time Sobol' indices. 
			\item GSIs - generalized Sobol' indices. 
			\item LMSIs - limited-memory Sobol' indices. 
			\item $i = 1,\dots,p$ denotes the parameter index.
		\end{tablenotes} 
	\end{threeparttable}
	\renewcommand{\arraystretch}{1}  
\end{table} 

\subsubsection{Time-varying Sobol' indices} 
\label{TimevaryingSobolindices}

\par \noindent  The following formulations attempt to account for changes in parameter influence over time.

\par \noindent {\bf Pointwise-in-time Sobol' indices (PTSIs)}: Consider a model $f$ with time-varying output $y(t)$ on the interval $[0, T]$ for $T > 0$, i.e., $y(t) = f(t;\theta)$ for $t \in  [0, T].$ The main effect PTSI of $f$ corresponding to $\theta_i$ at time $t$ for $t \in [0, T]$ is 
\begin{equation} 
	\mathscr{S}_i (f; t) = \dfrac{\mathbb{V}_{\theta_i} (\mathbb{E}_{\theta_{\sim i}} [f(t;\cdot) | \theta_i ])}{\mathbb{V}(f(t;\cdot))}
	\label{pwfosobol} 
\end{equation}
and the total effect PTSI at time $t$ is
\begin{equation} 
	\mathscr{T}_i (f;t) = \dfrac{\mathbb{E}_{\theta_{\sim i}}[\mathbb{V}_{\theta_i}(f(t;\cdot) |\theta_{\sim i})]}{\mathbb{V}(f(t;\cdot))}.
	\label{pwtesobol}
\end{equation} 

\par \noindent {\bf Generalized Sobol' indices (GSIs)}: The PTSIs ignore the variance history of the signal whereas GSIs take it into account by integrating the numerators and denominators of equations \eqref{pwfosobol} and \eqref{pwtesobol} \cite{Alexanderian2020}. The main effect GSI over the interval $[0, t]$ for $t \in [0, T]$  is
\begin{equation} 
	\mathscr{S}_i(f; [0,t]) = \dfrac{ \displaystyle\int_{0}^{t} \mathbb{V}_{\theta_i} (\mathbb{E}_{\theta_{\sim i}} [f(\tau; \cdot )|\theta_i]) \ \mathrm{d} \tau}{ \displaystyle\int_0^t \mathbb{V} (f(\tau; \cdot)) \ \mathrm{d}\tau }. 
	\label{tvfosobol}
\end{equation} 
Similarly, the total effect GSI of $\theta_i$ is
\begin{equation} 
	\mathscr{T}_i (f; [0, t]) = \dfrac{ \displaystyle\int_0^t \mathbb{E}_{\theta_{\sim i }} [ \mathbb{V}_{\theta_i} (f(\tau;\cdot ) | \theta_{\sim i })] \ \mathrm{d}\tau}{ \displaystyle\int_0^t \mathbb{V}(f(\tau; \cdot )) \ \mathrm{d}\tau}. 
	\label{tvtesobol} 
\end{equation}  

\par
\begin{figure}[!t]
	\centering 
	\includegraphics[]{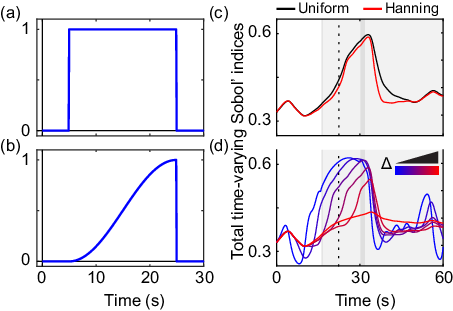}
	\caption{Moving integration window, $\Delta$, for the limited-memory Sobol' indices using parameter $H_s$ as an example. Valsalva maneuver (VM) phases are represented as alternating gray (I and III) and light gray (II and IV) boxes. Phase II is divided into early and late stages (vertical dotted black line). (a) Uniform moving window. (b) Half-Hanning moving window. (c) Comparison of uniform (black) versus Hanning (red) windows of width $\Delta = 15$ s. The Hanning window shows a steeper decline in parameter influence in phase IV. (d) Comparison of varying window lengths for increasing $\Delta = [5, 10, 15, 20]$ s. Pointwise-in-time Sobol' indices (blue) and generalized Sobol' indices (red) are also plotted.}
	\label{windowfig}
\end{figure}

 \noindent {\bf Limited-memory Sobol' indices (LMSIs)}: By integrating over [0, t], GSIs compute the parameter influence up to time $t$, which can average the signal and miss parameter contributions during fast, transient disturbances. Hence, we propose a new technique, {\em limited-memory} Sobol' indices, to analyze parameter influence as the QoI responds to these disturbances. To do so, we introduce a moving integration window of width $\Delta$. Thus, we have the interval $[t - \Delta, t]$ for $t \in [\Delta, T]$. The window shape and magnitude of $\Delta$ is problem-dependent. For $t \in [\Delta, T]$, the main effect LMSI of $\theta_i$ is
\begin{equation} 
	\mathscr{S}_i(f; [t-\Delta, t] ) = \dfrac{ \displaystyle\int_{t-\Delta}^t \mathbb{V}_{\theta_i} (\mathbb{E}_{\theta_{\sim i}}[f(\tau; \cdot )|\theta_i]) \ w(\tau) \ \mathrm{d} \tau}{ \displaystyle\int_{t-\Delta}^t  \mathbb{V}(f(\tau; \cdot )) \ w(\tau) \ \mathrm{d}\tau },
	\label{windfosobol}
\end{equation}  
where $w(t)$ is the weight determined by the window of choice. Figure \ref{windowfig} gives different options for $w(t)$, which are discussed in detail in Section \ref{Integrationwindow}. Similarly, the total effect LMSI is 
\begin{equation} 
	\mathscr{T}_i (f; [t-\Delta,t]) = \dfrac{ \displaystyle\int_{t-\Delta}^t  \mathbb{E}_{\theta_{\sim i }} [ \mathbb{V}_{\theta_i} (f(\tau; \cdot ) | \theta_{\sim i })] \ w(\tau) \ \mathrm{d}\tau}{ \displaystyle\int_{t-\Delta}^t  \mathbb{V} (f(\tau; \cdot )) \ w(\tau) \ \mathrm{d}\tau}.
	\label{windtesobol}
\end{equation} 

\par There are many factors in choosing the appropriate window type, shape, and width. We let the physiology guide our window selection. Since the blood pressure and the change in blood pressure affect the baroreflex modulation of heart rate, recent time points contribute to the current heart rate and future points do not. To accommodate these features, we chose to use a trailing half-Hanning integration window (Figure \ref{windowfig}b) for $\Delta = 15$ s to coincide with the typical length of the VM as
\begin{equation} 
w(t) = 0.5 (1 - \cos(\pi t / T)). 
\end{equation} 

\subsection{GSA computation}
\label{GSAcomputation}

\par \noindent In our numerical experiments, we use Monte Carlo integration to estimate the SIs of all model parameters ($p = 23$) except the VM start and end times, $t_s$ and $t_e$, which are extracted from data. Specifically, the SIs are calculated using the procedure outlined in Saltelli et al. \cite{Saltelli2010} (page 262, Table 2) that computes the expectations and variances. To test for convergence, we computed $10^3(p+2) = 25,000$, $10^4(p+2) = 250,000$, and $10^5(p+2) = 2,500,000$ function evaluations, which produced similar results \cite{Hart2018,Link2018}. For the reduced models, we use 25,000 evaluations.  To approximate the integrals in equations \eqref{tvfosobol}-\eqref{windtesobol}, we used the trapezoid quadrature scheme. 

\par We compare the performance of each of the four methods discussed in Section \ref{Globalsensitivityanalysis}. The scalar SIs are calculated with respect to the Euclidean norm of the residual $\mathbf{r}$ (equation \eqref{normr}), that is, $\mathscr{S}_i (||\mathbf{r}||_2)$ and $\mathscr{T}_i (||\mathbf{r}||_2)$ for $\mathbf{r}$. Using $||\mathbf{r}||_2$ as the scalar model output gives a indication of the average sensitivity of $\mathbf{r}$ to the parameters at steady-state. For the time-varying $\mathbf{r}$, we simultaneously compute the PTSIs, GSIs, and LMSIs.

\subsection{Sensitivity ranking} 
\label{Sensitivityranking}

\par
\begin{figure*}[!t]
	\centering
	\includegraphics[]{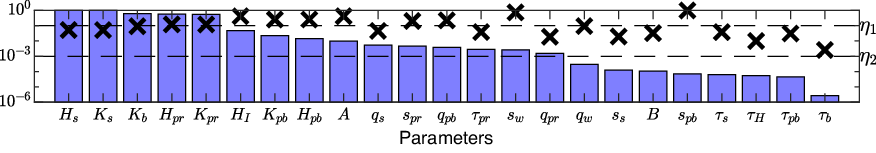}
	\caption{Log-scale relative scalar sensitivity analysis ranking with respect to $||\mathbf{r}||_2$. Influence thresholds $\eta_1 = 10^{-1}$ and $\eta_2 = 10^{-3}$ are indicated (horizontal dashed lines).  Local analysis results (x's) reproduced from \cite{Randall2019_JAP} are scaled from zero to one by the maximum sensitivity value. Global analysis results (bars) are computed using the scalar total effect Sobol' indices ($\mathscr{T}_i (||\mathbf{r}||_2)$, equation \eqref{scalartesobol}) and scaled by the maximum sensitivity value for comparison with the local results.}
	\label{compareSA} 
\end{figure*}

 \noindent To identify noninfluential parameters, we rank $\mathscr{T}_i (||\mathbf{r}||_2)$ and group them into most, moderately, and least influential.  For moderately influential, $\eta_1 = 10^{-1}$ is assigned based on the clear separation in parameter influence (Figure \ref{compareSA}). For least influential, $\eta_2 = 10^{-3}$ is assigned in accordance with the LSA from our previous work \cite{Randall2019_JAP}. The scalar SI groups motivated the grouping used for the PTSIs, GSIs, and LMSIs. We define a parameter $\theta_i$ as {\em noninfluential} over the time series if 
\begin{equation} 
	\mathscr{T}_i(\mathbf{r}; \cdot ) < \eta_2 
	\label{noninfluentialcriterion}
\end{equation}\
for the entire interval. 

\subsection{Model reduction} 
\label{Modelreductionmethods}

\par \noindent Using the sensitivity ranking, we develop steps for a model reduction methodology informed by the LMSIs. The steps are: 
\begin{enumerate}
	\item Compute the LMSIs for all parameters to be considered. 
	\item Determine if each $\mathscr{T}_i (f;[t-\Delta,t])$ for $t \in [\Delta,T]$ satisfies equation \eqref{noninfluentialcriterion} and make a set of noninfluential parameters, $\theta_{NI}$. 
	\item Analyze the parameters in $\theta_{NI}$ and determine if it is possible to remove the equations associated with each parameter. Choose parameter $\theta_k$ that has the least influence. Remove the components of the model associated with that parameter and restructure the model. Some changes to the model equations must occur simultaneously, which we will exemplify in the next section. Note that this step is problem-specific and inherently changes the mathematical structure of the model. There may be instances where removing model components could be detrimental. Also, there may be instances where there are multiple options available, creating branches of models. Then, analyze the parameters in $\theta_{NI}$ of each branch separately.  
	\item Recalibrate parameters to obtain new parameter vector $\theta^\star$ for newly generated model $f^\star$. $f^\star$ joins the set $\mathscr{M}$. 
	\item Repeat steps 1-4 on $f^\star$. Iterate until the equations associated with the noninfluential parameters in $\theta_{NI}$ are not algebraically removable. Fix all remaining parameters in $\theta_{NI}$ at their nominal values. 
\end{enumerate}
This process generates a set of models $\mathscr{M} = \{m_0, m_1, \dots, m_M\}$ for $m_0$ the original model and $m_i$ the reduced models for $i = 1, \dots, M$. 

\subsection{Model selection}
\label{Modelselectionmethods}

\par \noindent To compare model performance between the original and reduced models, we compute the AICc and BIC indices, statistical measures that balance the models' fit to the data (goodness of fit) and the model complexity (the number of estimated parameters) \cite{Qureshi2019}. The AICc index, derived from a frequentist perspective, overestimates the model fit to the data, whereas the BIC index, derived from a Bayesian perspective, underestimates it. These indices are commonly calculated for model selection and, if in agreement, provide confidence that the model selected balances over- and underfitting \cite{Burnham2002}. We assume that the residual errors are independent and identically distributed with mean zero and finite variance. By predicting the maximum likelihood estimate, or equivalently minimizing the least squares error $J$ (equation \eqref{J}), we compute 
\begin{align} 
	\text{AICc} &= N \log \Bigg(\dfrac{J}{N}\Bigg) + 2(\hat{p}+2) \Bigg( \dfrac{N}{N - (\hat{p}+2) - 1} \Bigg) \quad \text{and} \label{AICc} \\
	\text{BIC} &= N \log \Bigg(\dfrac{J}{N}\Bigg) + (\hat{p}+2)\log(N) \label{BIC}, 
\end{align} 
where $N$ is the number of data points, $\hat{p}$ the dimension of the optimized parameter subset $\hat{\mathbf{\theta}}$  \cite{Burnham2002}. To compare the models, we report the relative index of model $m_i$ from the minimal model, that is, 
\begin{equation}
	\text{AICc}_{ri} = e^{(\text{AICc}_{m} - \text{AICc}_i)/2} \quad \text{and} \quad \text{BIC}_{ri} = e^{(\text{BIC}_{m} - \text{BIC}_{i})/2},
	\label{AICcr_BICcr}
\end{equation}
where $\text{AICc}_m$ and $\text{BIC}_m$ are the minimum AICc and BIC values, and $\text{AICc}_i$ and $\text{BIC}_i$ are the corresponding values for model $m_i$. 

\par Though this statistical technique is useful when determining the goodness of fit to data, there are predicted model outputs (e.g., $T_{pb}$ and $T_s$) of clinical importance that cannot be measured without costly and invasive procedures. Therefore, we must also assess the model performance qualitatively. To do so, we compare the reduced models to $m_0$, assuming that $m_0$ produces the true signal. We employ the metric 
\begin{equation}
	Q = |\max \big(T_{s}^{m_0}(t)\big) - \max \big(T_{s}^{m_i}(t) \big)|,
	\label{Tsmetric}
\end{equation} 
where $T_{s}^{m_0}(t)$ is the baroreflex sympathetic tone $T_s(t)$ from $m_0$ and $T_{s}^{m_i}(t)$ is $T_s(t)$ from the reduced model $m_i$. 

%%%%%%%%%%%%%%%%%%%%%%%%%%%%%%%%%%%%%%%%%%%%%%%%%%%%%%%%%%%%%%%%%%%%%%%%%%%%%%%%%%%%%%%%%%%%%%%%%%%%%%%%%%%%%%%%%%%%%%%%%

\begin{figure}[!t]
	\centering
	\includegraphics[]{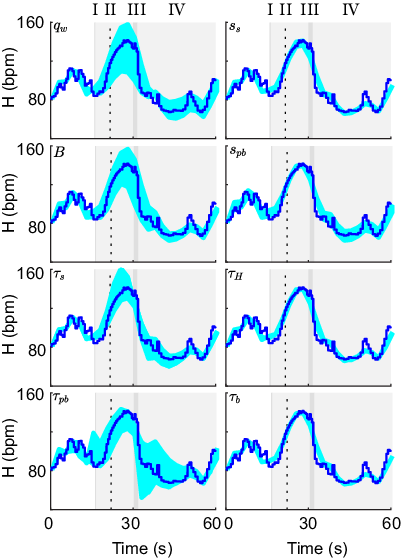}
	\caption{Noninfluential parameters determined by the total effect scalar Sobol' indices in Figure \ref{compareSA}. The model was evaluated at 10 equidistant parameter values within the physiological range of each parameter given in Table \ref{parameters} (cyan) plotted with the heart rate data (blue).  The VM phases are represented as alternating gray (I and III) and light gray (II and IV) boxes. Phase II is divided into early and late parts (vertical dotted black line). } 
	\label{oneatatime} 
\end{figure} 

\section{Results}
\label{Results}

\par
\begin{figure*}[!t]
	\centering 
	\includegraphics[]{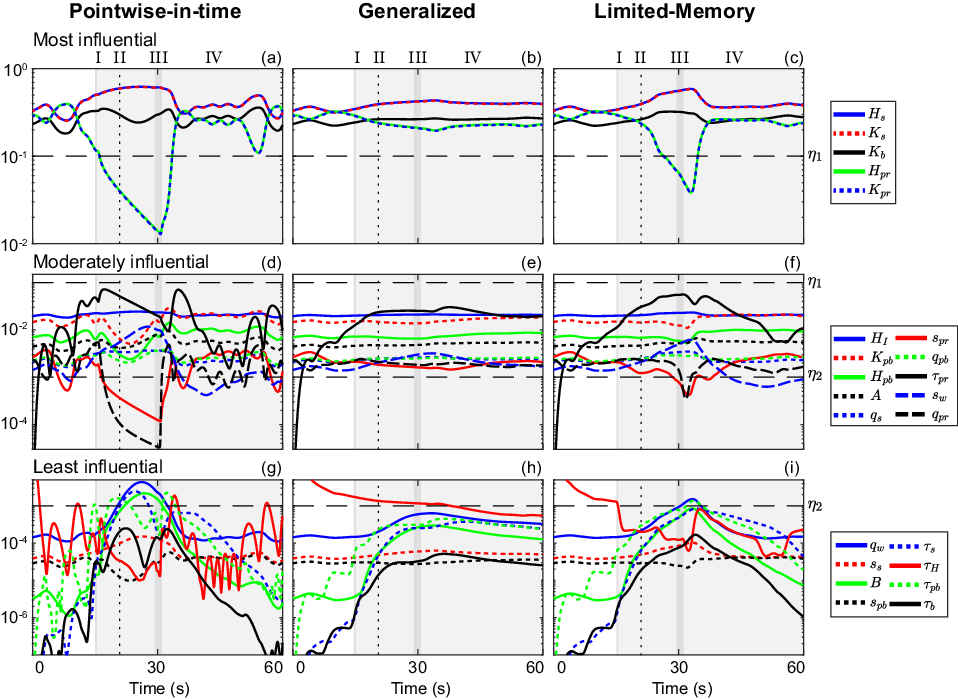}
	\caption{Time-varying total effect Sobol' indices (SIs) for the most (top row), moderately (middle row), and least (bottom row) influential parameters determined by the scalar SIs in Figure \ref{compareSA} for the pointwise-in-time (a, d, g), generalized (b, e, h), and limited-memory (c, f, i) SIs. Thresholds $\eta_1 = 10^{-1}$ and $\eta_2 = 10^{-3}$ are indicated with horizontal dashed black lines. The VM phases are represented as alternating gray (I and III) and light gray (II and IV) boxes. Phase II is divided into early and late parts (vertical dotted black line). Results are plotted on a logarithmic scale for the $y$-axis.}
	\label{timevaryingindices}
\end{figure*}

 \noindent This section discusses the results regarding the scalar SIs (Figure \ref{compareSA}) and the time-varying PTSIs, GSIs, and LMSIs (Figure \ref{timevaryingindices}). The latter are computed using a moving integration window of width $\Delta$ and several window widths are compared (Figure \ref{windowfig}d). For each analysis, the parameters are divided into three influence groups: most ($> \eta_1$), moderately (between $\eta_1$ and $\eta_2$), and least ($< \eta_2$) influential. Using the LMSIs, we inform a model reduction protocol. Lastly, we select the best model by computing the $\text{AICc}_{r}$ and $\text{BIC}_{r}$ (equation \eqref{AICcr_BICcr}) values in Table \ref{statistics} and by examining the predicted neural tones $T_{pb}$ and $T_s$ to give physiological predictions (Figure \ref{RMcomparison}).

\subsection{Scalar Sobol' indices}
\label{ResultsScalarSIs}

\par \noindent The total effect scalar SIs are calculated with respect to $||\mathbf{r}||_2$, that is, $\mathscr{T}_i (||\mathbf{r}||_2)$ and plotted with LSA results reproduced from \cite{Randall2019_JAP} (Figure \ref{compareSA}).  The five most influential parameters, $\{H_s, K_s, K_b, H_{pr}, K_{pr}\}$, are associated with RSA and $T_s$ and affect $H$ during and after the VM. In comparison, the LSA determined that the parameters $s_{pb}$, $s_w$, $A$, and $H_I$ were the most influential. The subset of least influential parameters is 
\begin{equation} 
	\theta_{NI}^S = \{ q_w, s_s, B, s_{pb}, \tau_s, \tau_H, \tau_{pb}, \tau_b\},
	\label{NIsca}
\end{equation} 
where the superscript $S$ denotes scalar SIs. Figure \ref{oneatatime} displays $H$ when varying each parameter at 10 equidistant points between their upper and lower bounds (given in Table \ref{parameters}) and all other parameters are held fixed. Parameters $s_s$, $s_{pb}$, $\tau_H$, and $\tau_b$ have the least effect visually on the model output. The small effect of changing $s_s$ and $s_{pb}$ is most likely due to their narrow distributions. In \cite{Randall2019_JAP}, these parameters are calculated {\it a priori} and do not vary much among the subjects. Parameters $\tau_H$ and $\tau_b$ have the least effect on the residual for both analyses.

\subsection{Pointwise-in-time Sobol' indices}
\label{ResultsPTSIs}

\par \noindent Figure \ref{timevaryingindices} shows results for the most (Figure \ref{timevaryingindices}a), moderately (Figure \ref{timevaryingindices}d), and least (Figure \ref{timevaryingindices}g) influential parameters for the PTSIs, which display rapid fluctuations. The most influential parameters, $H_s$, $K_s$, $K_b$, $H_{pr}$, and $K_{pr}$ (Figure \ref{timevaryingindices}a), correspond to the scalar SIs. Our results show that the index for $K_b$ does not change significantly during the time interval, which is expected since the baroreceptors are always active. Parameters $K_s$ and $H_s$ are most influential due sympathetic activation after the breath hold. Finally, $H_{pr}$ and $K_{pr}$ decrease in influence below $\eta_1$ after the breath hold due to parasympathetic withdrawal. 

\par Parameters $s_{s}$ and $s_{pb}$ remain below $\eta_2$ for all time. Though all other time-scales rise above $\eta_2$, $\tau_b$ surprisingly also remains below $\eta_2$. These insensitive parameters are prime candidates for removal. Overall, the PTSIs designate almost all parameters as at least moderately influential, which agree with the LSA results. However, these results do not provide much additional information that is not already given in the scalar SIs. 

\subsection{Generalized Sobol' indices}
\label{ResultsGSIs}

\par \noindent As shown in Figures \ref{timevaryingindices}b, e, and h, the GSIs smooth the signals, placing much emphasis on the time-dependence and averaging the signal over extended periods of baseline activity. They miss potential transient changes in parameter influence on the model output. PTSIs and GSIs provide extremes for analyzing time-varying signals as shown in Figure \ref{windowfig}d (PTSI in blue and GSI in red). These results suggest that there is a need for a method that incorporates both the transient nature of the PTSIs and the smoothing capabilities of the GSIs. 

\par The GSI for $\tau_H$ (Figure \ref{timevaryingindices}h) fluctuates significantly. $\tau_H$ is moderately influential at the beginning but decreases in influence after the onset of the VM. Hence, $\tau_H$ may play a role in establishing steady-state behavior from the initial conditions and then decrease in influence once steady-state is achieved. As shown in Figure \ref{timevaryingindices}h, the least influential subset of parameters determined by the GSIs is 
\begin{equation} 
	\theta_{NI}^{G} = \{q_w, s_s, B, s_{pb}, \tau_s, \tau_{pb}, \tau_b \}, 
	\label{NIG}
\end{equation} 
where the superscript $G$ denotes generalized SIs.

\subsection{Limited-memory Sobol' indices}
\label{ResultsLMSIs}

\par \noindent For the LMSIs, it is important to choose a window that is short enough so as to retain important features but long enough to be easily interpretable. Since the VM typically lasts 15 s, we compare the output from several window widths for $\Delta = 5, 10, 15,$ and $20$ s (Figure \ref{windowfig}d). For smaller values of $\Delta$, the LMSI tends toward the PTSI whereas for larger values of $\Delta$, the LMSI tends toward the GSI. To remain consistent with the VM itself, we chose a window of 15 s. 

\par Fluctuations in parameter influence correspond to different control mechanisms that activate and deactivate during the VM. The LMSIs remain relatively constant before the VM (Figures \ref{timevaryingindices}c, f, and i). We observe increases in influence of parameters associated with $T_s$ during the VM that decrease afterward. This behavior coincides with the activation and deactivation of $T_s$ during the VM. Note that $q_w$ and $\tau_{pb}$ become moderately influential during the VM while the subset
\begin{equation} 
\theta_{NI}^{LM} = \{ s_s, B, s_{pb}, \tau_s, \tau_b \},
\label{NILM}
\end{equation}
where superscript $LM$ denotes limited-memory, remains below $\eta_2$. $\tau_b$ has consistently been the least influential throughout all analyses. Surprisingly, $B$ and $\tau_s$ also remain below $\eta_2$ for all $t$. 

\begin{table}[!t]
	\centering 
	\begin{threeparttable} 
		\renewcommand\arraystretch{1}
		\caption{Noninfluential (NI) parameters.}
		\begin{tabular}{cccc}
			\hline
			{\bf Parameters} & $\theta_{NI}^{S}$ \eqref{NIsca} & $\theta_{NI}^{G}$ \eqref{NIG} & $\theta_{NI}^{LM}$ \eqref{NILM} \\ 
			\hline 
			\hline 
			$q_w$ &$\checkmark$ & $\checkmark$  & \\
			$s_s$ & $\checkmark$ & $\checkmark$ & $\checkmark$ \\
			$B$ & $\checkmark$ &  $\checkmark$ & $\checkmark$ \\
			$s_{pb}$ & $\checkmark$ &  $\checkmark$ & $\checkmark$ \\ 
			$\tau_s$  & $\checkmark$ & $\checkmark$ &  $\checkmark$ \\
			$\tau_H$ & $\checkmark$ &  & \\
			$\tau_{pb}$ & $\checkmark$  & $\checkmark$ &  \\ 
			$\tau_b$ & $\checkmark$ & $\checkmark$  & $\checkmark$ \\
			\hline 
		\end{tabular} 
		\label{noninfluentialparams}
		\begin{tablenotes} 
			\small
			\item S - scalar. G - generalized. LM - limited memory. 
		\end{tablenotes} 
	\end{threeparttable} 
\end{table} 

\par In summary, the scalar SIs and LSA provide clear parameter influence rankings. However, the time-varying techniques elucidate {\em when} a parameter becomes influential. Our results show that the GSIs have a clear benefit over the PTSIs, incorporating the signal history, but miss the effects of transient disturbances. We conclude that the LMSIs illustrate parameter influence traces that correspond to the modeled physiological phenomena. Table \ref{noninfluentialparams} summarizes the subsets of noninfluential parameters from the scalar SIs (equation \eqref{NIsca}), GSIs (equation \eqref{NIG}), and LMSIs (equation \eqref{NILM}). The PTSIs were not included as their results were inconclusive. The LMSIs determined the smallest subset with 5 noninfluential parameters considered for removal.  

\subsection{Model reduction} 
\label{Modelreduction}

\par \noindent We ``remove" a parameter from consideration by fixing it at its nominal value or analytically excising equations associated with it. In Section \ref{fixedparameters}, we consider the parameters in equation \eqref{NILM} to fix at their nominal values. In Sections \ref{removingtaub} and \ref{removingB}, we remove equations associated with noninfluential parameters and develop a suite of models upon which we can perform statistics. 

\begin{table}[!t]
	\centering 
	\renewcommand{\arraystretch}{1}
	\begin{threeparttable} 
		\caption{Estimated parameter values.}
		\begin{tabular}{cccccc}
			\hline
			{\bf Parameters} & $m_0$ & $m_1$ & $m_2$  & $m_3$ & $m_4$ \\ 
			\hline
			\hline
			$B$              & 0.43	 & 0.43   & 0.58   & 0$^\dagger$ & 1$^\dagger$\\
			$\tau_{pb}$  & 4.27	  & 4.27  & 4.05    & 3.85 & 4.62 \\
			$\tau_{pr}$   & 2.71   & 2.71   & 2.61    &	2.68  & 2.58 \\
			$H_{pb}$      & 0.43  & 0.43  &	0.43    & 0.42 & 0.43\\
			$H_{pr}$       & 0.48  & 0.48  & 0.46    & 0.50 & 0.39\\
			$H_s$           & 0.28  & 0.28  &  0.32   &	0.20 & 0.47\\
			\hline
		\end{tabular} 
		\label{estparams}
		\begin{tablenotes} 
			\small
			\item $^\dagger$ held constant and excluded from estimation. 
		\end{tablenotes} 
	\end{threeparttable} 
\end{table}

\subsubsection{Fixed parameters}
\label{fixedparameters}

\par \noindent Parameters $s_{pb}$ and $s_s$ (equations \eqref{Gpb} and \eqref{Gs}) are computed {\it a priori} assuming 80\% of the baseline heart rate is controlled by $T_{pb}$ and 20\% by $T_s$ \cite{Randall2019_JAP}. The coefficient of variation (SD/mean) for both is $\sim$0.1\%, which implies a very narrow dispersion. In \cite{Randall2019_MathBiosci}, we show that $\tau_s$ (equation \eqref{Ts}) is nonlinearly related to $D_s$, which can lead to instability. Therefore, we restricted the bounds of $\tau_s$ to $\pm 50\%$ of the nominal value (Table \ref{parameters}) to ensure model stability. After fixing these parameters at their nominal values, we recalculated the LMSIs with 25,000 samples, which produced similar results to Figure \ref{timevaryingindices}i with $\tau_b$ and $B$ still in $\theta_{NI}^{LM}$ (results not shown). This process produces the reduced model, $m_1$, for which $s_{pb}$, $s_s$, and $\tau_s$ are constant. Table \ref{estparams} lists parameters for $m_0$ and $m_1$. Note that both models have the same estimated parameters. Therefore, from here on, we compare further reduced models to $m_1$. $\tau_b$ and $B$ are removed and the equations impacted by these parameters are modified as described in the following sections. 

\subsubsection{Removing $\tau_b$}
\label{removingtaub}

\par \noindent Since  $\tau_b$ (equations \eqref{ebc} and \eqref{eba}) can be small with negligible impact, we rearrange equation \eqref{ebc} as 
\begin{equation} 
	\tau_b \fder{\varepsilon_{bc}(t)}{t} + \varepsilon_{bc}(t) = K_b G_{bc} (t). 
	\label{eps}
\end{equation} 
By letting $\tau_b = 0$, equations \eqref{ebc}  and \eqref{eba} become $\varepsilon^*_{bc}(t) = K_b G_{bc} (t)$ and $\varepsilon^*_{ba}(t) = K_b G_{ba}(t)$, respectively. From here on, an asterisk denotes the reduced system. Note that the dimension of the state space has reduced by two. Substituting $\varepsilon^*_{bc}$ and $\varepsilon^*_{ba}$ into equation \eqref{n} gives  
\begin{align} 
	n(t) &= B(G_{bc} (t) - \varepsilon^*_{bc}(t)) + (1 - B) ( G_{ba} (t) - \varepsilon^*_{ba}(t)) \nonumber \\ 
	&= (1 - K_b) n^*(t), 
	\label{newn}
\end{align} 
where
\begin{align}
	n^*(t) = BG_{bc}  (t) + (1 - B) G_{ba} (t). 
	\label{nstar} 
\end{align}
By substituting equation \eqref{newn} into equation \eqref{Gpb}, we obtain 
\begin{equation} 
	G^*_{pb}(t) = \dfrac{1}{1 + e^{-q_{pb} (n(t) - s_{pb})}} = \dfrac{1}{1 + e^{-q^*_{pb} (n^*(t) - s^*_{pb})}}
	\label{newGpb},
\end{equation}
and define new parameter values  
\begin{equation}
	q^*_{pb} = q_{pb} (1 - K_b) \quad \text{and}  \quad s^*_{pb}  =  \dfrac{s_{pb}}{1-K_b}. \label{qpb_ss_star}
\end{equation} 
$G_s(t)$ also reduces to $G_s^*(t)$ with new parameters $q^*_{s}$ and $s^*_{s}$ (equation \eqref{Gsstar}). Note that $K_b$ is used to calculate $q^*_{pb}$, $s^*_{pb}$, $q^*_s$, and $s^*_s$, and hence, the dimension of $\Omega_p$ is reduced by one.

\par The reduced system consists of 4 states and 23 parameters
\begin{equation}
	\fder{\mathbf{x}^*(t)}{t} = f^* (t, \mathbf{x}^*(t), \mathbf{x}^*(t - D_s); \mathbf{\theta}^*), 
	\label{reducedtaub}
\end{equation}
where $\mathbf{x}^*(t) = [T_{pb}(t), T_{pr}(t), T_s(t), H(t)]^T \in \mathbb{R}^4$ and 
\begin{multline} 
	\theta^* = [A, B, K_{pb}, K_{pr}, K_s, \tau_{pb}, \tau_{pr}, \tau_{s}, \tau_H, q_w, q_{pb}^*, q_{pr}, q_s^*, \\ s_w, s_{pb}^*, s_{pr}, s_s^*, H_I, H_{pb}, H_{pr}, H_s, t_s, t_e]^T \in \mathbb{R}^{23}.
\end{multline} 
Note that $\theta^*$ does not contain $K_b$. The full set of equations of $f^*$ are in \ref{appendix:fulleqs}. We refer to this reduced model as $m_2$. 

\par We perform subset selection using the structured correlation analysis described in \ref{appendix:subsetselection} to determine a subset of parameters to optimize for $m_2$ and obtain 
\begin{equation}
	\hat{\theta}^{m_2} = [B, \tau_{pb}, \tau_{pr}, H_{pb}, H_{pr}, H_s]^T. 
\end{equation}
Table \ref{estparams} lists the optimized parameters for $m_2$. We perform GSA on $m_2$, holding $s_{pb}$, $s_s$, and $\tau_s$ fixed, and observe that $B$ is the only parameter below $\eta_2$ for all time $t$ (results not shown). \\

\subsubsection{Removing $B$}
\label{removingB} 

\par \noindent The necessity of delineating the aortic and carotid regions has been explored in two previous studies \cite{Kosinski2018,Randall2019_JAP} and, surprisingly, $B$ is flagged for removal. Note that when $B = 0$ in equation \eqref{nstar}, the aortic strain solely influences the efferent response, and when $B = 1$, the carotid. It is unknown how the these signals are integrated in the medulla and not clear whether it is sufficient to model only one of them. We observe that the LMSI of $B$ is small initially, increases as the VM progresses, and decreases after the breath hold (Figure \ref{timevaryingindices}i). This produces two models branching from $m_2$:  $m_3$ where $B = 0$ and $n^*_2(t) = G_{ba} (t)$ and $m_4$ where $B = 1$ and	$n^*_3(t) =  G_{bc} (t)$. These are systems of ODEs and DDEs with 4 states and 22 parameters of the form in equation \eqref{reducedtaub}.  \ref{appendix:fulleqs} contains more details for the construction of $m_3$ and $m_4$. Table \ref{estparams} lists the estimated parameters for $m_3$ and $m_4$. We perform GSA on each, which results in all parameters above $\eta_2$ at some time point (results not shown). Hence, there are no parameters remaining in the noninfluential subset.

\begin{figure}[!t]
	\centering
	\includegraphics*[]{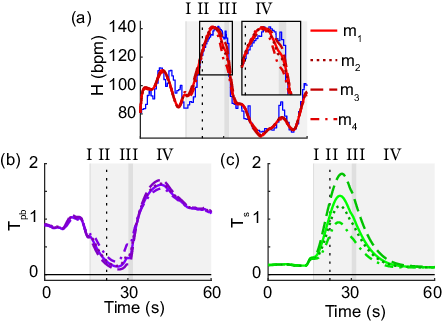}
	\caption{Plots of the reduced model 1 ($m_1$, solid), 2 ($m_2$, dotted), 3 ($m_3$, dashed), and 4 ($m_4$, dash-dotted). The original model $m_0$ has been omitted since it has the same estimated parameters as $m_1$ (Table \ref{estparams}). The Valsalva maneuver (VM) phases are represented as alternating gray (I and III) and light gray (II and IV) boxes. Phase II is divided into early and late stages (vertical dotted black line). (a) Model fits (red) to heart rate data (blue). Insert shows a zoom in late phase II and phase III. (b) Model predictions of baroreflex-mediated parasympathetic ($T_{pb}$, purple) for the full and reduced models. (c) Model predictions of baroreflex-mediated sympathetic ($T_s$, green) for the original and reduced models.} 
	\label{RMcomparison}
\end{figure} 

\par In summary, the model reduction methodology has produced 5 possible models (Table \ref{estparams}): $m_0$, the original model; $m_1$, the original model with $s_{pb}$, $s_s$, and $\tau_s$ fixed;  $m_2$ reduced model removing $\tau_b$; $m_3$ reduced model removing $\tau_b$ and setting $B = 0$; and $m_4$ reduced model removing $\tau_b$ and setting $B = 1$. Figure \ref{RMcomparison} plots the model fits to the data as well as their respective neural tones. 

\subsection{Model selection} 
\label{Modelselection}

\par \noindent To determine which model best fits the data, we perform model selection quantitatively by computing AICc$_r$ and BIC$_r$ (equation \eqref{AICcr_BICcr}) and qualitatively by comparing $T_{pb}$ and $T_s$ of the reduced models to $m_0$. The latter is important since our goal is not only to fit the data but also to predict the neural tones that cannot be measured {\it in vivo}. 

\subsubsection{Goodness of fit analysis}

\par \noindent Table \ref{statistics} shows that $m_2$ has the greatest $\text{AICc}_{r}$ and $\text{BIC}_{r}$ values. This is surprising since $m_2$ estimates more model parameters than both $m_3$ and $m_4$. This analysis suggests that modeling both the regions is necessary to predict heart rate. 

\subsubsection{Qualitative assessment} 

\par \noindent Figure \ref{RMcomparison} shows $H$, $T_{pb}$, and $T_s$ for all of the models considered: $m_1$ (solid curve), $m_2$ (dotted curve), $m_3$ (dashed curve) and $m_4$ (dash-dotted curve). The model fits to heart rate data (Figure \ref{RMcomparison}a) are all similar, which is expected. The least squares costs of the fits are of the same magnitude (Table \ref{statistics}). Interestingly, $m_3$ is the only model that is able to fit phase III of the data (Figure \ref{RMcomparison}a insert), suggesting from a qualitative standpoint that $m_3$ captures the most features of the heart rate data. Figure \ref{RMcomparison}b displays the predicted $T_{pb}$ trace for the models, which are all similar. Figure \ref{RMcomparison}c shows that $T_s$, exhibits the greatest deviation from $m_0$. Thus, we use $T_s$ to compare the reduced models to $m_0$ via the $Q$ value  (equation \eqref{Tsmetric}). Table \ref{statistics} compiles these metrics and $m_2$ has the lowest $Q$ value, suggesting that both regions must be included. Hence, we conclude that $m_2$ is the best model for the biological questions investigated here since the $\text{AICc}_{r}$ and $\text{BIC}_{r}$ values are the greatest while the qualitative metric $Q$ was the lowest. 

\begin{table}[!t]
	\centering 
	\renewcommand{\arraystretch}{.625}
	\begin{threeparttable} 
		\caption{Statistical analysis for model selection.}
		\begin{tabular}{cccccc}
			\hline
			{\bf Model} & {\bf Cost} & {\bf Pars} & \multicolumn{2}{c}{{\bf Quantitative}} & {\bf Qualitative}  \\ 
			\cline{4-5}
			& $J$ ($10^{-3}$)  & $p$ & $\text{AICc}_{r}$ & $\text{BIC}_{r}$  & $Q$ \\ 
			& \eqref{J}  & & \eqref{AICcr_BICcr}  & \eqref{AICcr_BICcr} &  \eqref{Tsmetric} \\
			\hline 
			\hline 
			$m_0$ & 4.64 & 6 & 0.83 & 0.83 & \\ 
			$m_1$ & 4.64 & 6 & 0.83 & 0.83 & 0 \\
			$m_2$ & 4.63 &  6 & 1 & 1 & 0.17 \\
			$m_3$ & 4.97 & 5 & 2.5e-4 & 3.6e-4 & 0.41 \\
			$m_4$ & 4.97 &  5 & 2.5e-5 & 3.6e-5 & 0.48\\  
			\hline 
		\end{tabular} 
		\label{statistics}
		\begin{tablenotes} 
			\small
			\item Pars - number of parameters. AICc - Akaike Information Criterion with correction. BIC - Bayesian Information Criterion.  
		\end{tablenotes} 
	\end{threeparttable} 
\end{table}  

%%%%%%%%%%%%%%%%%%%%%%%%%%%%%%%%%%%%%%%%%%%%%%%%%%%%%%%%%%%%%%%%%%%%%%%%%%%%%%%%%%%%%%%%%

\section{Discussion}
\label{Discussion} 

\par \noindent This study performs model reduction and selection using GSA on a cardiovascular model predicting heart rate ($H$) in response to the VM. Computation for the scalar SIs \cite{Saltelli2010} was done with respect to $||\mathbf{r}||_2$ while the time-varying PTSIs \cite{Alexanderian2012}, GSIs \cite{Alexanderian2020}, and LMSIs are with respect to $\mathbf{r}$. A novel component of this study is the introduction of the LMSIs using a moving integration window $\Delta$ motivated by the transient VM. The scalar SIs determined a ranking of parameter influence, from which we created three groups: most, moderately, and least influential. Additionally, the LMSIs informed a model reduction protocol that generated five models: $m_0$, $m_1$, $m_2$, $m_3$, and $m_4$. We analyzed the performance of these models quantitatively, calculating the $\text{AICc}_{r}$ and $\text{BIC}_{r}$ (equation \eqref{AICcr_BICcr}) values, and qualitatively, comparing the model predicted signals to the original model $m_0$, $Q$ (equation \eqref{Tsmetric}). From this analysis, $m_2$ is the best model, suggesting that modeling both the aortic and carotid baroreceptor regions are  necessary to predict heart rate and neural tones. 

\subsection{Local versus global sensitivity analysis}
\label{LSAvsGSA}

\par \noindent This study focused on using GSA to analyze the influence of the parameters on the model output. These methods are computationally expensive, whereas LSA methods may suffice. In \cite{Randall2019_JAP}, we performed LSA on the neural model, reproduced in Figure \ref{compareSA}. LSA is useful in its  (i) relative ease in computation and (ii) ability to calculate time-varying sensitivities. In steady-state with nominal parameters close to their optimal value, LSA is useful in ranking parameter influence \cite{Ellwein2008,Marquis2018,Olufsen2013}. The results from the LSA are similar to the ranking of the scalar SIs.  Since every parameter in the LSA was above $\eta_2 = 10^{-3}$, every parameter is influential. However, this is a snapshot of the model sensitivity at one instance in $\Omega_p$. The benefit of GSA is that it explores the entirety of $\Omega_p$ and incorporates parameter interactions. Though there are some differences, the parameter rankings for both the LSA and scalar SIs are similar overall, which could be due to the {\it a priori} calculation of nominal parameter values. Others have also found agreement in the calculation of the local and global parameter influences \cite{Link2018}. 

\subsection{Time-varying GSA} 
\label{TimevaryingGSA_discussion}

\par \noindent Though there are other methods that explore $\Omega_p$ \cite{Olsen2019,Sobol2010}, we used SIs due to their broad applications. Here, we introduce LMSIs that provide a more informative ranking than scalar SIs during a transient disturbance from baseline. They smooth the oscillatory PTSIs and incorporate the variance history preserved by the GSIs. The LMSIs retain some modulation, providing distinct changes in parameter influence rankings before, during, and after the VM. Lastly, they can be used for virtually any modeling effort analyzing parameter influence over time. 

\par Not every problem requires time-varying GSA, which depends on the questions asked and the QoI. Scalar SIs are appropriate if the QoI is a scalar, in steady-state, at a particular time point, or periodic. If the objective is to understand parameter influence during state transition, time-varying GSA may provide rich information. One such study is that of Calvo et al. \cite{Calvo2018}, which calculated SIs for the parameters of a cardiovascular model studying head-up tilt at rest and during the tilt. The state transition as the tilt occurs is of great clinical interest, and hence, this study can benefit from using LMSIs to measure changes in parameter influence. Another is the study by Sumner et al. \cite{Sumner2012} that claims to analyze time-dependent parameter influence with SIs. However, the QoI for this model is a state evaluated at time point $t = 60$, which is a scalar value. If other time points are of interest, either GSIs or LMSIs can be used to quantify the parameter influence over time. 

\subsection{Integration window} 
\label{Integrationwindow}

\par \noindent Moving windows have been used in signal processing for decades and recently in graphical sensitivity analysis \cite{Iooss2015,Storlie2008}, typically smoothing signals. There are several considerations when choosing an appropriate $\Delta$ for analysis of a biological problem: shape, type, and width. Shape refers to the functional form providing the weights for the window. There are many potential window shapes, but uniform (Figure \ref{windowfig}a) and Hanning (Figure \ref{windowfig}b) windows are common. A uniform window applies equal weight to each $t_j$ whereas a half-Hanning places more emphasis on the most recent time points. Figure \ref{windowfig}c compares the traces from both a uniform (black) and a Hanning (red) shape, though the differences are negligible. For the physiological application examined here, the only reasonable window type is a trailing window. Width refers to the time interval over which the window applies. We chose $\Delta = 15$ s to correspond with the VM. We strongly suggest allowing the features of the system studied to dictate choice of window. 
 
\subsection{Model reduction} 
\label{Modelreduction_discussion}

\par \noindent Due to the overall model complexity, understanding the biological implications of the results and parameter interactions can be difficult. Therefore, model reduction can simplify these interactions and still retain its predictive power. Many model reduction techniques exist from engineering and control theory \cite{Besselink2013}, aiming to reduce large numbers of state variables with many nonlinearities by attempting to mitigate the extent that the input parameters affect the output. Our method using GSA to inform an analytical model reduction uses this idea to make appropriate choices for the exclusion of certain model components, unlike methods that solely approximate input-output relationships without considering other predicted model quantities \cite{Snowden2018}. To our knowledge, there are no previous studies that inform a model reduction based on time-varying GSA methods for physiological models. For biological systems with scalar QoIs, there has been a methodology proposed for model reduction via GSA by Marino et al. \cite{Marino2008}; however, a statistical analysis of a group of reduced models was not conducted. Furthermore, we propose our GSA-informed model reduction methodology as an alternative approach to the balanced truncation method \cite{Snowden2018}.  

\par Though we acknowledge that SIs may be impractical for large ODE systems (e.g.  pharmacokinetics models), we suggest a multi-level approach. One can use GSA measures that are simpler to compute but possibly less informative first to identify a set of unimportant parameters and then perform a more comprehensive analysis on the remaining parameters \cite{Hart2019}. 

\subsection{Model selection}
\label{ModelselectionDis}

\par \noindent Previous studies have modeled baroreceptor stimulation of only the carotid region \cite{LeRolle2008,Lu2001}, though to our knowledge there are no studies solely modeling the aortic baroreceptors. To our knowledge, no previous studies have performed a model selection protocol for cardiovascular models in response to the VM. To consider the effect of reducing $m_0$ on the predicted quantities, we combine quantitative and qualitative approaches to select whether delineating between carotid and aortic regions is necessary, and, if not, which pathway should be modeled. We perform our analysis on our set of models $\mathscr{M} = \{ m_0 m_1,, m_2, m_3, m_4 \}$, and results show that $m_2$ has the greatest AICc$_r$ and BIC$_r$ values and the lowest $Q$ value. Ultimately, we conclude that $m_2$ is the best model both to fit the heart data and predict the neural tones, and therefore, both regions are necessary to model the VM, agreeing with previous studies \cite{Kosinski2018,Randall2019_JAP}.

\subsection{Limitations} 
\label{Limitations}

\par \noindent The GSA method of choice is dependent on the model formulation and the QoI. GSA may not be feasible for computationally intensive models. GSA results will differ based on the choice of QoI. For our GSA, SIs assume that the model parameters are statistically independent, which may not be reasonable. Furthermore, analytical model reduction may be impractical for large systems. In this case, setting noninfluential parameters to their nominal values may be more reasonable. Lastly, using AICc$_r$ and BIC$_r$ values is dependent on available data. If a different heart data set had been used, the outcome of the model selection protocol could be different. However, we conducted this analysis on three representative subjects and achieved similar results (not shown). 

\section{Conclusions} 
\label{Conclusions} 

\par \noindent In this study, we introduced a model reduction and selection methodology informed by global sensitivity analysis on a model predicting heart rate and autonomic function in response to the Valsalva maneuver. We developed the novel limited-memory Sobol' indices that account for both the variance history and transient dynamics. Furthermore, with extensive numerical experiments, we simplified the model while retaining important physiological characteristics. We conclude that modeling both the aortic and carotid regions is necessary to achieve the appropriate dynamics of the Valsalva maneuver. 

\section{Acknowledgments}

\par \noindent We would like to thank Dr. Pierre Gremaud for his help developing and analyzing the limited-memory Sobol' indices. 

\section{Funding} 

\par \noindent This study was supported in part by the National Science Foundation under awards NSF/DMS 1246991 and NSF/DMS 1557761 and by the National Institute of Health HL139813.

\appendix

\setcounter{table}{0}
\renewcommand{\thetable}{A.\arabic{table}}

\setcounter{equation}{0}
\renewcommand{\theequation}{A.\arabic{equation}}

\section{Model equations}
\label{appendix:fulleqs}

\par \noindent This appendix lists the equations for the original model, $m_0$, and the reduced models, $m_1$, $m_2$, $m_3$, and $m_4$, given in Section \ref{Methods}. 

\par \noindent {\bf Original model.} $m_0$ (Section \ref{Modeldevelopment}) has 6 states and 25 parameters of the form in equation \eqref{summary}. The system $f$ \cite{Randall2019_JAP} consists of the following set of equations:
\begin{equation}
	\fder{\varepsilon_{bc}(t)}{t} = \dfrac{1}{\tau_b} ( -\varepsilon_{bc}(t) + K_b G_{bc} (t)), \label{ebc} 
\end{equation} 
\begin{equation} 
	\fder{\varepsilon_{ba}(t)}{t} = \dfrac{1}{\tau_b} ( -\varepsilon_{ba}(t) + K_b G_{ba} (t)), \label{eba}
\end{equation} 
\begin{equation} 
	\fder{T_{pb}(t)}{t} = \frac{1}{\tau_{pb}} (-T_{pb}(t) + K_{pb} G_{pb}(t)), \label{Tpb} 
\end{equation} 
\begin{equation} 
	\fder{T_{s}(t)}{t} = \frac{1}{\tau_{s}} (-T_{s}(t - D_s) + K_{s} G_{s}(t)), \label{Ts}
\end{equation} 
\begin{equation} 
	\fder{T_{pr}(t)}{t} = \frac{1}{\tau_{pr}} (-T_{pr}(t) + K_{pr} G_{pr}(t)), \label{Tpr}
\end{equation} 
\begin{equation} 
	\fder{H(t)}{t} = \frac{1}{\tau_H}(-H(t) + \tilde{H}(t)), \label{H}
\end{equation} 
\begin{equation} 
	P_{t}(t) = \left\{ 
		\begin{array}{ll} 
		\text{ITP}(t) & t_s \leq t \leq t_e \\ 
		\dfrac{R_M - R_m}{\bar{R}_I - \bar{R}_E} R(t) + (R_m - \bar{R}_E) & \text{otherwise}
		\end{array}, 
		\right. 	\label{Pth}  
\end{equation} 
\begin{equation} 
	P_c(t) = \text{SBP}(t), \label{Pc}
\end{equation} 
\begin{equation} 
	P_a(t) = \text{SBP}(t) - P_{t}(t),  \label{Pa}
\end{equation} 
\begin{equation}
	G_{bc}(t) = 1 - \sqrt{\dfrac{1 + e^{-q_w (P_c(t) - s_w)}}{A + e^{-q_w (P_c(t) - s_w)}}}, \label{ewc}
\end{equation} 
\begin{equation} 
	G_{ba}(t)  = 1 - \sqrt{\dfrac{1 + e^{-q_w (P_a(t) - s_w)}}{A + e^{-q_w (P_a(t) - s_w)}}}, \label{ewa}
\end{equation} 
\begin{equation} 
	n(t) = B (G_{bc} (t)-\varepsilon_{bc}(t)) + (1 - B) ( G_{ba} (t)-\varepsilon_{ba}(t)), 
\end{equation} 
\begin{equation} 
	G_{pb}(t) = \dfrac{1}{1 + e^{-q_{pb} (n(t) - s_{pb})}}, \label{Gpb}
\end{equation} 
\begin{equation} 
	G_{s}(t)  = \dfrac{1}{1 + e^{q_{s} (n(t) - s_{s})}}, \label{Gs}
\end{equation} 
\begin{equation} 
	G_{pr} (t) = \dfrac{1}{1 + e^{-q_{pr} (P_{t}(t) - s_{pr})}}, \label{Gpr}
\end{equation} 
\begin{equation} 
	\tilde{H}(t) = H_I (1 - H_{pb} T_{pb}(t) + H_{pr} T_{pr}(t) + H_s T_s(t)). \label{Htilde} 
\end{equation}
	
\par \noindent {\bf Reduced models.} The reduced model for $m_1$ has the same equations as $m_0$. $m_2$ (Section \ref{Modelreduction}) has 4 states and 23 parameters and is of the form in equation \eqref{reducedtaub}. The system $f^*$ consists of equations \eqref{Tpr} - \eqref{ewa}, equation \eqref{Gpr}, and the equations:
	
\begin{equation}
	\fder{T_{pb}(t)}{t} = \frac{1}{\tau_{pb}} (-T_{pb}(t) + K_{pb} G_{pb}^{*}(t)),
\end{equation} 
\begin{equation} 
	\fder{T_{s}(t)}{t} = \frac{1}{\tau_{s}} (-T_{s}(t - D_s) + K_{s} G_{s}^{*}(t)),
\end{equation} 
\begin{equation} 
	n^*(t) = B G_{bc} (t) + (1 - B)G_{ba} (t),
	\label{nstar2} 
\end{equation} 
\begin{equation} 
	G_{pb}^{*}(t) = \dfrac{1}{1 + e^{-q^*_{pb} (n^*(t) - s^*_{pb})}},
\end{equation} 
\begin{equation}  
	G_{s}^{*}(t)  = \dfrac{1}{1 + e^{q^*_{s} (n^*(t) - s^*_{s})}}, \label{Gsstar}
\end{equation} 
For reduced model $m_3$, $B = 0$, equation \eqref{nstar2} becomes $n^*_2(t) = G_{ba} (t)$, and \eqref{Pc} and \eqref{ewc} are eliminated. For $m_4$, $B = 1$, equation \eqref{nstar2} becomes $n^*_3(t) = G_{bc}(t)$, \eqref{Pth}, \eqref{Pa}, and \eqref{ewa} are eliminated. 

\section{Initial conditions} 
\label{appendix:ICs}

\setcounter{equation}{0}
\renewcommand{\theequation}{B.\arabic{equation}}

\par \noindent  This appendix contains calculations of the initial conditions (ICs) for $f$ in equation \eqref{summary} computed to ensure the model is in steady-state. The ICs take into account the baseline systolic blood pressure $\bar{P}$, thoracic pressure $\bar{P}_{t}$, and heart rate $\bar{H}$, that is, $\bar{P}_c = \bar{P}$ and $\bar{P}_a = \bar{P} - \bar{P}_{t}$. We let $\bar{\cdot}$ denote steady-state. The ICs for $\varepsilon_{bc}$ and $\varepsilon_{ba}$ (equations \eqref{ebc} and \eqref{eba}) are 
\begin{align}
	\bar{\varepsilon}_{bc} &= K_b \bar{G}_{bc}  = K_b \Bigg(1 - \sqrt{\dfrac{1 + e^{-q_w (\bar{P}_c - s_w)}}{A + e^{-q_w (\bar{P}_c - s_w)}}}\Bigg) \quad \text{and} \label{ewcbar}\\
	\bar{\varepsilon}_{ba} &= K_b \bar{G}_{ba} = K_b \Bigg(1 - \sqrt{\dfrac{1 + e^{-q_w (\bar{P}_a - s_w)}}{A + e^{-q_w (\bar{P}_a - s_w)}}}\Bigg). \label{ewabar} 
\end{align}
The ICs $\bar{T}_{pb} = 0.8$ and $\bar{T}_s = 0.2$ are based on the assumption that 80\% of baroreflex contribution to $H$ is due to the parasympathetic tone and 20\% due to sympathetic \cite{Randall2019_JAP}. The IC for the RSA-mediated parasympathetic tone is  
\begin{equation} 
\bar{T}_{pr} = K_{pr} \bar{G}_{pr} = \dfrac{K_{pr}}{1 + e^{-q_{pr}(\bar{P}_{t} - s_{pr})}}. \label{Tprbar} \\
\end{equation} 
The IC for the model output heart rate is $\bar{H}$. 

\section{Subset selection and parameter estimation} 
\label{appendix:subsetselection}

\setcounter{equation}{0}
\renewcommand{\theequation}{C.\arabic{equation}}

\par \noindent This appendix provides details concerning parameter subset selection and estimation. We determine a subset of parameters $\hat{\theta}$ to optimize using structured correlation analysis \cite{Olufsen2013,Pope2009}. The $i^\text{th}$ column of a sensitivity matrix ($\mathbf{S}$) of the model residual $\mathbf{r}$ with respect to the logarithm of the parameters at time $t_j$ is 
\begin{equation} 
	S_{ij} = \fpartial{\mathbf{r}(t_j)}{\log \theta_i} = \fpartial{H(t_j;\theta)}{\theta_i} \dfrac{\theta_i}{H_d(t_j)} 
\end{equation} 
for $H$ the model output heart rate and $H_d$ the data. Note that this analysis is local. To determine possible pairwise correlations between influential parameters \cite{Olufsen2013}, we calculate a correlation matrix $\mathbf{c}$ from a covariance matrix $\mathbf{C} = (\mathbf{\hat{S}}^T \mathbf{\hat{S}})^{-1}$ as 
\begin{equation} 
	c_{ij} = \dfrac{C_{ij}}{\sqrt{C_{ii}C_{jj}}},
\end{equation} 
where $\hat{\mathbf{S}}$ is a matrix with the columns of $\mathbf{S}$ corresponding to the parameters in $\hat{\theta}$. $\mathbf{c}$ is symmetric with $|c_{ii}| =1$ and $|c_{ij}| \leq 1$. We assigned a threshold of 0.9 for correlated parameters. 

\par We removed some parameters from consideration {\it a priori}, such as all $q_x$ and $s_x$. Optimizing these parameters can force the model to produce results that are not physiological \cite{Randall2019_JAP}. We also excluded $t_s$ and $t_e$ from consideration as they were determined directly from the data and are naturally very influential.

\par From these methods, we obtain the identifiable subset for the original model as 
\begin{equation}
	\hat{\theta} = [B,\tau_{pb}, \tau_{pr}, H_{pb}, H_{pr}, H_s]^T.
	\label{thetahat} 
\end{equation}
More details can be found in \cite{Randall2019_JAP}.  Given $\mathbf{r}$ (equation \ref{residual}), we fit $H$ to data by minimizing the least squares cost
\begin{equation}
	J(\hat{\theta}) = \mathbf{r}(t)^T \mathbf{r}(t) + \Bigg( \dfrac{ \max\limits_j H_m(t_j; \theta) - \max\limits_j H_d(t_j)}{\max\limits_j H_d(t_j)} \Bigg)^2.
	\label{J} 
\end{equation}

%%%%%%%%%%%%%%%%%%%%%%%%%%%%%%%%%%%%%%%%%%%%%%%%%%%%%%%%%%%%%%%%%%%%%%%%%%%%%%%%%%%%%

\bibliographystyle{elsarticle-num-mod} 
%\bibliography{MyEndnoteLibrary}


\end{document}